\documentclass{article}

\usepackage[square,numbers]{natbib}
\usepackage[preprint]{neurips_2019}

\usepackage[utf8]{inputenc} % allow utf-8 input
\usepackage[T1]{fontenc}    % use 8-bit T1 fonts
\usepackage{hyperref}       % hyperlinks
\usepackage{url}            % simple URL typesetting
\usepackage{booktabs}       % professional-quality tables
\usepackage{amsfonts}       % blackboard math symbols
\usepackage{nicefrac}       % compact symbols for 1/2, etc.
\usepackage{microtype}      % microtypography
\usepackage{multirow}

\usepackage{fourier} 
\usepackage{array}
\usepackage{makecell}

\usepackage{graphicx}
\newcommand{\dcell}[2]{#1$\pm$#2}
\newcommand{\dcellbf}[2]{\textbf{#1}$\pm$\textbf{#2}}
\newcommand{\dcellit}[2]{\textit{#1}$\pm$\textit{#2}}

\title{The International Workshop on Osteoarthritis Imaging Knee MRI Segmentation Challenge: A Multi-Institute Evaluation and Analysis Framework on a Standardized Dataset}

\author{%
  Arjun D. Desai \\
  Stanford University\\
  \texttt{arjundd@stanford.edu} \\
   \And
   Francesco Caliva \\
   University of California, \\ San Francisco \\
   \AND
   Claudia Iriondo \\
   University of California, \\ San Francisco \\
   \And
   Naji Khosravan \\
   University of Central Florida \\
   \And
   Aliasghar Mortazi \\
   University of Central Florida \\
   \And
   Sachin Jambawalikar \\
   Columbia University \\
   \And
   Drew Torigian \\
   University of Pennsylvania \\
   \And
   Jutta Ellermann \\
   University of Minnesota \\
   \And
   Mehmet Ak\c{c}akaya \\
   University of Minnesota \\
   \And
   Ulas Bagci \\
   University of Central Florida \\
   \And
   Radhika Tibrewala \\
   University of California, \\ San Francisco \\
   \And
   Io Flament \\
   University of California, \\ San Francisco \\
   \And
   Matthew O'Brien \\
   University of California, \\ San Francisco \\
   \And
   Sharmila Majumdar \\
   University of California, \\ San Francisco \\
   \And
   Mathias Perslev \\
   University of Copenhagen \\
   \And
   Akshay Pai \\
   University of Copenhagen \\
   \And
   Christian Igel \\
   University of Copenhagen \\
   \And
   Erik B. Dam \\
   University of Copenhagen \\
   \And
   Sibaji Gaj \\
   Cleveland Clinic \\
   \And
   Mingrui Yang \\
   Cleveland Clinic \\
   \And
   Kunio Nakamura \\
   Cleveland Clinic \\
   \And
   Xiaojuan Li \\
   Cleveland Clinic \\
   \And
   Cem M. Deniz \\
   New York University,\\ Langone Health \\
   \And
   Vladimir Juras \\
   Medical University of Vienna\\
   \And
   Ravinder Regatte \\
   New York University,\\ Langone Health \\
   \And
   Garry E. Gold \\
   Stanford University \\
   \And
   Brian A. Hargreaves \\
   Stanford University \\
   \And
   Valentina Pedoia \\
   University of California,\\ San Francisco\\
   \And
   Akshay S. Chaudhari \\
   Stanford University \\
}

\begin{document}

\maketitle
\pagebreak
\begin{abstract}
\textbf{Purpose}: To organize a knee MRI segmentation challenge for characterizing the semantic and clinical efficacy of automatic segmentation methods relevant for monitoring osteoarthritis progression. \\

\textbf{Methods}: A dataset partition of consisting of 3D knee MRI from 88 subjects at two timepoints, with ground-truth articular (femoral, tibial, patellar) cartilage and meniscus segmentations was standardized. Challenge submissions and a majority vote ensemble were evaluated using Dice score, average symmetric surface distance, volumetric overlap error, and coefficient of variation on a hold-out-test-set. Similarities in network segmentations were evaluated using pairwise Dice correlations. Articular cartilage thickness was computed per-scan and longitudinally. Correlation between thickness error and segmentation metrics was measured using Pearson’s coefficient. Two empirical upper bounds for ensemble performance were computed using combinations of model outputs that consolidated true positives and true negatives.\\

\textbf{Results}: Six teams ($T_1$-$T_6$) submitted entries for the challenge. No significant differences were observed across all segmentation metrics for all tissues (p=1.0) among the four top-performing networks ($T_2$, $T_3$, $T_4$, $T_6$). Dice correlations between network pairs were high (>0.85). Per-scan thickness errors were negligible among networks $T_1$-$T_4$ (p=0.99) and longitudinal changes showed minimal bias (<0.03mm). Low correlations ($\rho$<0.41) were observed between segmentation metrics and thickness error. The majority vote ensemble was comparable to top performing networks (p=1.0). Empirical upper bound performances were similar for both combinations (p=1.0). \\

\textbf{Conclusion}: Diverse networks learned to segment the knee similarly where high segmentation accuracy did not correlate to cartilage thickness accuracy. Voting ensembles did not exceed individual network performance but may help regularize individual models. 

\end{abstract}

\section{Introduction}
Osteoarthritis (OA) affects over 30 million adults and is the leading cause of chronic disability in the United States \cite{zhang2010epidemiology, cross2014global}. The current gold-standard for OA detection is radiography, which can only detect late-stage OA changes due to its lack of sensitivity to soft-tissue degeneration \cite{kellgren1958osteo, guermazi2011radiography}. Magnetic resonance imaging (MRI) provides excellent soft tissue contrast and recent studies have shown that morphological and compositional changes in the articular cartilage and meniscus are potential biomarkers for early OA \cite{chaudhari2019rapid}. Accurately measuring such tissue properties relies on high-quality tissue segmentations, the gold-standard for which is a manual approach. However, manual annotations can be time-consuming and prone to inter-reader variations, leading interest in automated cartilage and meniscal MRI segmentation techniques.

Convolutional neural networks (CNNs) have shown great potential due to their high accuracies and ease-of-use. While increases in computational power and open-source frameworks have made CNN development pervasive, they have also contributed to complications in their practical use. As the performance of deep-learning models is data-dependent, different algorithms must be evaluated on the same subset of data for fair comparison. However, networks are often evaluated on different datasets or different partitions of the same dataset (i.e. different holdout splits), making it challenging to compare segmentation methods.

Data standardization for semantic segmentation of knee MRI has been explored in previous organizational challenges, such as MICCAI’s SKI10 challenge, which provided 1.5T and 3.0T MRI data for segmenting bone and cartilage in the femoral and tibial condyles \cite{heimann2010segmentation}. While this challenge was instrumental in creating one of the first standardized datasets for knee segmentation, curated datasets that standardize scan contrasts and field strengths and that include additional tissue compartments could be useful for future research. Recent studies have shown that changes in meniscus and patellar cartilage morphology are correlated to OA progression \cite{draper2006cartilage, emmanuel2016quantitative}. Larger initiatives such as the Osteoarthritis Initiative (OAI) have standardized protocols for imaging these soft-tissues and have publicly shared expert-annotated segmentations for a subset of scans. However, different automatic segmentation methods have used different data partitions, making it difficult to accurately compare these methods \cite{desai2019technical, norman2018use, ambellan2019automated, gaj2019automated, panfilov2019improving, chaudhari2019utility}. Different data partitions also preclude properly combining and evaluating predictions from multiple CNNs through model ensembles, which may be superior to a single top-performing model in medical imaging tasks \cite{pan2019improving}.

In this work, we describe the organization and present results from the 2019 International Workshop on Osteoarthritis Imaging (IWOAI) knee segmentation challenge that introduced a standardized partition for data in the OAI. We present a framework to compare and evaluate the performance of challenge submission entries for segmenting articular (femoral, tibial, and patellar) cartilage and the meniscus. We also characterize the extent to which traditional segmentation metrics correlate with clinically relevant endpoints such as cartilage thickness. Finally, we evaluate the potential for utilizing an ensemble of networks and provide an exploratory analysis of how to empirically quantify upper bounds on segmentation performance.

\section{Methods}
\subsection{Data}
\subsubsection{Image Dataset}
Data for this study originated from the Osteoarthritis Initiative (OAI)\footnote{http://oai.epi-ucsf.org}, a longitudinal study of OA progression. 3D sagittal double echo steady state (DESS) and corresponding segmentation masks for femoral, tibial, and patellar cartilage, and meniscus generated by Stryker Imorphics (Manchester, United Kingdom) were used in this study (DESS MRI parameters: FOV=14cm, resolution=0.31mm$\times$0.46mm$\times$0.70mm, TE/TR=5/16ms, number of slices = 160) \cite{williams2010measurement}. The dataset comprised of 88 subjects with Kellgren-Lawrence (KL) grade OA between 1 and 4 scanned at two timepoints (baseline, 1 year), resulting in 176 annotated 3D DESS volumes \cite{kellgren1957radiological}.  Overall, a total of 28,160 segmented slices with four different tissue classes were included in this challenge. The 88 subjects were split into cohorts of 60 subjects for training, 14 for validation, and 14 for testing, resulting in 120, 28, and 28 volumes for training, validation, and testing, respectively, with approximately equal distributions of KL grade, BMI, and sex among all three groups (see Table \ref{tbl:data-distribution}). 

\begin{table}
  \caption{Subject distribution (mean $\pm$ standard deviation) in training, validation and testing datasets. Age, Kellgren-Lawrence Grade (KLG) and BMI were calculated for subjects at the first timepoint. Asterisk ($\ast$) indicates significance (p<0.001).}
  \label{tbl:data-distribution}
  \centering
  \begin{tabular}{ccccccccc}
    \toprule
    & & & & & \multicolumn{4}{c}{\textbf{\% KLG}} \\
    \cmidrule(r){6-9}
    & & \textbf{Count} & \textbf{Age (range)} & \textbf{BMI} & \textbf{1} & \textbf{2} & \textbf{3} & \textbf{4} \\
    \midrule
    \multirow{3}{*}{\textbf{Train}} & Male & 31 & 58$\pm$10 (45-78)$^\ast$ & 31$\pm$4 & 2 & 16 & 31 & 3 \\
    & Female & 29 & 58$\pm$9 (46-78) & 33$\pm$5 & 1 & 20 & 26 & 3 \\
    & Total & 60 & 58$\pm$9 (45-78) & 32$\pm$5 & 3 & 36 & 57 & 5 \\
    \midrule
    \multirow{3}{*}{\textbf{Validation}} & Male & 5 & 68$\pm$8 (52-72)& 29$\pm$1 & 4 & 11 & 18 & 11 \\
    & Female & 9 & 64$\pm$4 (57-71) & 28$\pm$4 & 0 & 18 & 39 & 0 \\
    & Total & 14 & 65$\pm$6 (52-72) & 29$\pm$3 & 4 & 29 & 57 & 11 \\
    \midrule
    \multirow{3}{*}{\textbf{Test}} & Male & 9 & 73$\pm$4 (65-78) & 31$\pm$4 & 0 & 25 & 29 & 7 \\
    & Female & 5 & 66$\pm$9 (49-76) & 31$\pm$4 & 0 & 7 & 29 & 4 \\
    & Total & 14 & 71$\pm$7 (49-78) & 31$\pm$4 & 4 & 32 & 57 & 11 \\
    \bottomrule
  \end{tabular}
\end{table}

\subsubsection{Distribution}
Deidentified training and validation sets, which included labeled masks as reference segmentations were shared with 29 researchers who requested access through the IWOAI website.  All participants were allowed to use training data from other sources and perform data augmentation. One week prior to the challenge, participants were provided with the test dataset, which consisted of scans without reference segmentations. All participants were asked to submit multi-label binarized masks for each scan in the test dataset along with an abstract with detailed CNN reporting categories, similar to the CLAIM guidelines (included in supplementary materials) \cite{mongan2020checklist}. 

\subsection{Challenge Entries}
Five teams participated in the challenge; while a sixth team submitted an entry following the challenge. Teams were numbered by submission time; ordering does not reflect performance ranking. A summary of submissions is shown in Table \ref{tbl:model-params}.

Team 1 ($T_1$) trained a modified multi-class 3D U-Net architecture, which used dilated convolutions at the bottleneck layer to increase the effective receptive field of the network \cite{deniz2018segmentation}. Volumes were zero-mean whitened (zero-mean, unit variance) for both training and inference.

Team 2 ($T_2$) implemented a cascaded ensemble of 3D and 2D variants of the V-Net architecture using dropout, mimicking a coarse-to-fine region annotation workflow \cite{milletari2016v, srivastava2014dropout, Iriondo2020}. In this pipeline, output segmentations from the 3D network were used to crop the scan volume around the centroid of each tissue, resulting in four cropped volumes, which were subsequently segmented by the 2D network. Data augmentation for both networks utilized randomly sampled intensity and geometric transforms. At test time, tissue probabilities from the 3D and 2D network were ensembled to produce the segmentation mask.

Team 3 ($T_3$) used a multi-planar sampling protocol to train a 2D U-Net architecture with batch normalization and nearest-neighbor upsampling \cite{perslev2019one}. During training, 2D slices were sampled at random across six fixed planes of random orientation through the volume. To augment the dataset, 2D slices were elastically deformed at random (with a probability of 0.33) and both original and left-/right mirrored volumes were used for training. During inference, volumes were segmented across the six planes and merged using a learned weighted linear combination of the six volumes.

Team 4 ($T_4$) designed a network modifying the DeeplabV3+ backbone to use dense connections at the bottleneck convolutional block \cite{chen2018encoder}. Atrous (dilated) spatial pyramid pooling with kernel sizes 1,6,12,18 was used at the end of DeeplabV3+ backbone to extract features at multiple scales. These features were merged, upsampled using nearest neighbor interpolation, and concatenated with corresponding encoder feature maps (i.e. skip connections).

Team 5 ($T_5$) implemented a 2D multi-view encoder-decoder that adaptively fused sagittal, axial, and coronal views to enforce high-level 3D semantic consistency \cite{mortazi2017cardiacnet}. Training data was augmented using translation, rotation, and scaling methods and subsequently preprocessed using anisotropic smoothing and histogram matching. The model was trained on the multi-class Z-loss \cite{de2016z}.

Team 6 ($T_6$) submitted a multi-class 2D U-Net to segment slices in the sagittal plane. The encoder consisted of 5 max pooling steps, after which the number of feature maps doubled. The network was trained on a multi-class soft-Dice objective with early stopping with patience of 8 epochs.

\begin{table}
  \caption{Summary of parameters used for training networks submitted by all participants. Participants with multiple associated networks ensembled outputs of different networks as part of their submission. WCE-- weighted cross entropy; $\eta$-- learning rate.}
  \label{tbl:model-params}
  \centering
  \begin{tabular}{ccccccc}
    \toprule
    \thead{Team} & \thead{Backbone\\(tissues segmented)} & \thead{Batch\\Size} & \thead{Optimizer\\(Parameters)} & \thead{Weight \\Initialization} & \thead{Activation} & \thead{Loss} \\
    \midrule
    1 & \makecell{3D U-Net\\(all)} & 1 & \makecell{RMSProp\\ ($\eta$=5e-5, $\alpha$=0.995)} & He & softmax & \makecell{WCE+\\ soft-Dice} \\
    \midrule
    \multirow{3}{*}{2} & \makecell{3D V-Net\\(patellar cartilage)} & 1 & \multirow{2}{*}{\makecell{Adam\\($\eta$=5e-5, $\beta_1$=0.9,\\$\beta_1$=0.999, $\epsilon=1e-8$}} & \multirow{3}{*}{Xavier} & sigmoid & soft-Dice \\
    & \makecell{3D V-Net\\(femoral cartilage,\\tibial cartilage,\\menisci)} & 1 & & & softmax & \makecell{weighted\\ soft-Dice} \\
    & \makecell{2D V-Net\\(per tissue)} & 8-16 & \makecell{Adam\\($\eta$=1e-4, $\beta_1$=0.9,\\$\beta_1$=0.999, $\epsilon=1e-8$} & & sigmoid & \makecell{soft-Dice} \\
    \midrule
    3 & \makecell{2D multi-planar\\U-Net\\(all)} & 16 & \makecell{Adam\\($\eta$=5e-5, $\beta_1$=0.9,\\$\beta_1$=0.999, $\epsilon=1e-8$} & \makecell{Glorot\\Uniform} & softmax & \makecell{Cross-\\entropy} \\
    \midrule
    4 & \makecell{2D DeeplabV3+\\Densenet\\(all)} & 4 & \makecell{Adam\\($\eta$=2e-4, $\beta_1$=0.5,\\$\beta_1$=0.999, $\epsilon=1e-8$} & \makecell{Glorot\\uniform} & softmax & soft-Dice \\
    \midrule
    5 & \makecell{2D encoder-decoder\\(all)} & 4 & \makecell{Adam\\($\eta$=1e-4, $\beta_1$=0.9,\\$\beta_1$=0.999, $\epsilon=1e-8$} & \makecell{Glorot\\uniform} & softmax & z-loss \\
    \midrule
    6 & \makecell{2D U-Net\\(all)} & 32 & \makecell{Adam\\($\eta$=1e-3, $\beta_1$=0.9,\\$\beta_1$=0.999, $\epsilon=1e-8$} & He & softmax & soft-Dice \\
    \bottomrule
  \end{tabular}
\end{table}

\subsection{Network Evaluation}
Networks were evaluated on the unreleased test set segmentations. Evaluation metrics for the challenge was limited to average Dice score (Dice, range:[0,1]) across for all tissues separately. Additionally, three other pixel-wise segmentation metrics were also computed – volumetric overlap error (VOE, range:[0,1]), root-mean-squared coefficient of variation (CV, range:[0,$\infty$]), and average symmetric surface distance (ASSD, range:[0,$\infty$)) in millimeters. For all metrics except Dice, a lower number indicated higher accuracy. 

To compute the similarity in segmentation results between different networks, the pairwise Dice correlations among test set predictions from all networks were calculated. The Dice correlation ($\rho_{Dice}$) between segmentation from network A ($f_A(x)$) and network B ($f_B(x)$) was defined as in Equation \ref{eq:dice}.

\begin{equation}
\label{eq:dice}
    	Dice\left(f_A(x),\ f_b\left(x\right)\right)=\ \frac{2\cdot f_A\left(x\right)\cdot f_B(x)}{f_A\left(x\right)+f_B(x)}\
\end{equation}

Depth-wise region of interest distribution (dROId) plots, which display 2D slice-wise Dice accuracies calculated over normalized knee sizes in the through-plane (left-right) dimension, were used to visualize differences in segmentation performance from the medial to lateral compartment \cite{desai2019technical}.

\subsection{Cartilage Thickness}
Cartilage thickness, a potential biomarker for knee OA progression, was also calculated for the three cartilage surfaces to assess the clinical efficacy and quality of automatic cartilage segmentations \cite{Iriondo2020, eckstein2015cartilage}. Cartilage thickness error was defined as the difference in the average thickness computed using the ground truth segmentation and the predicted segmentation. The correlation between pixel-wise segmentation metrics (Dice, VOE, CV, ASSD) and cartilage thickness error was measured by the Pearson’s correlation coefficient. Correlation values were described as very weak (Pearson-R=0-0.19), weak (Pearson-R=0.2-0.39), moderate (Pearson-R=0.4-0.59), strong (Pearson-R=0.6-0.79), and very strong (Pearson-R=0.8-1). As temporal change in cartilage thickness is the most common use of the thickness metrics, longitudinal thickness changes for all 14 subjects in the test set from baseline to 1 year were compared between the automated approaches across all networks and the manually annotated labels \cite{eckstein2015cartilage}. 

\begin{figure}
  \centering
  \includegraphics[width=8cm]{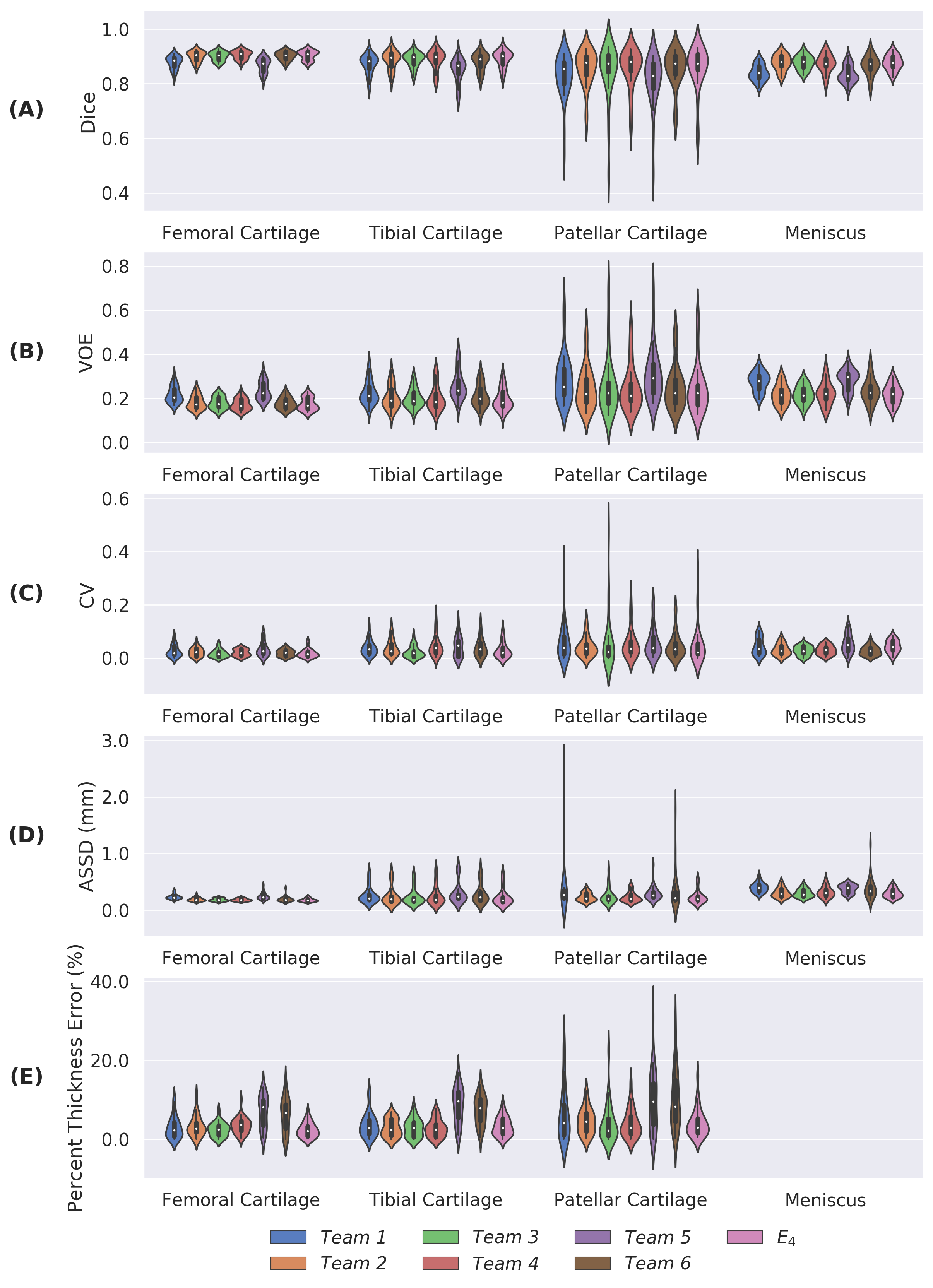}
  \caption{Performance summary of networks submitted to segmentation challenge and majority-vote ensemble ($E_4$) for all tissues as measured by Dice overlap (Dice), volumetric overlap error (VOE), coefficient of variation (CV), average symmetric surface distance (ASSD, in mm), and thickness error (mm). Network performances are indicated by violin plots, which overlay distributions over box plots. Longer plots indicate larger variance in network performance among scans. Thickness metrics were not calculated for meniscus.}
  \label{fig:perf-violin}
\end{figure}

\subsection{Network Ensembles}
In these experiments, we investigated how ensembles can be leveraged to improve prospective evaluation on unseen data (in the wild) and to empirically quantify performance bounds. We computed a majority-vote ensemble ($E_4$), which generated labels by selecting the super-majority (4 of 6) label across binarized segmentations submitted by each team. Performance was compared to that of individual networks using the metrics described above.

Additionally, CNN training protocols often recommend mitigating class imbalance by weighting the contribution of false positives and false negatives to regularize error rates. Loss functions like the Tversky loss tune the extent of false positive and false negative contributions to the final loss \cite{tversky1977features}. In an additional exploratory analysis, we utilized ensembles of submitted networks to empirically evaluate the upper bounds of segmentation sensitivity ($E_+^\ast$) and specificity ($E_-^\ast$) that is possible to achieve using a combination of the six networks. Ensembles $E_+^\ast$ and $E_-^\ast$ each consolidated true positives and true negatives from all networks, isolating errors in the segmentation to false negatives and false positives, respectively. Furthermore, using these upper-bound performance ensembles, we evaluated whether segmentation and cartilage thickness errors were lower for networks preferentially optimizing either for sensitivity or specificity (supplementary section \textit{Ensemble Upper Bounds}).

\subsection{Statistical Analysis}
Statistical comparisons were conducted using Kruskal-Wallis tests and corresponding Dunn post-hoc tests with Bonferroni correction ($\alpha$=0.05). All statistical analyses were performed using the SciPy (v1.1.0) library \cite{virtanen2020scipy}.

\section{Results}
\subsection{Data}
No significant difference was observed in the distribution of KL grades (p=0.51), BMI (p=0.33), and sex (p=0.41) among training, validation, and testing datasets (see Table \ref{tbl:data-distribution}). The age of subjects in the training set was significantly lower than in validation or testing sets (p<0.001). 

\begin{figure}
  \centering
  \includegraphics[width=8cm]{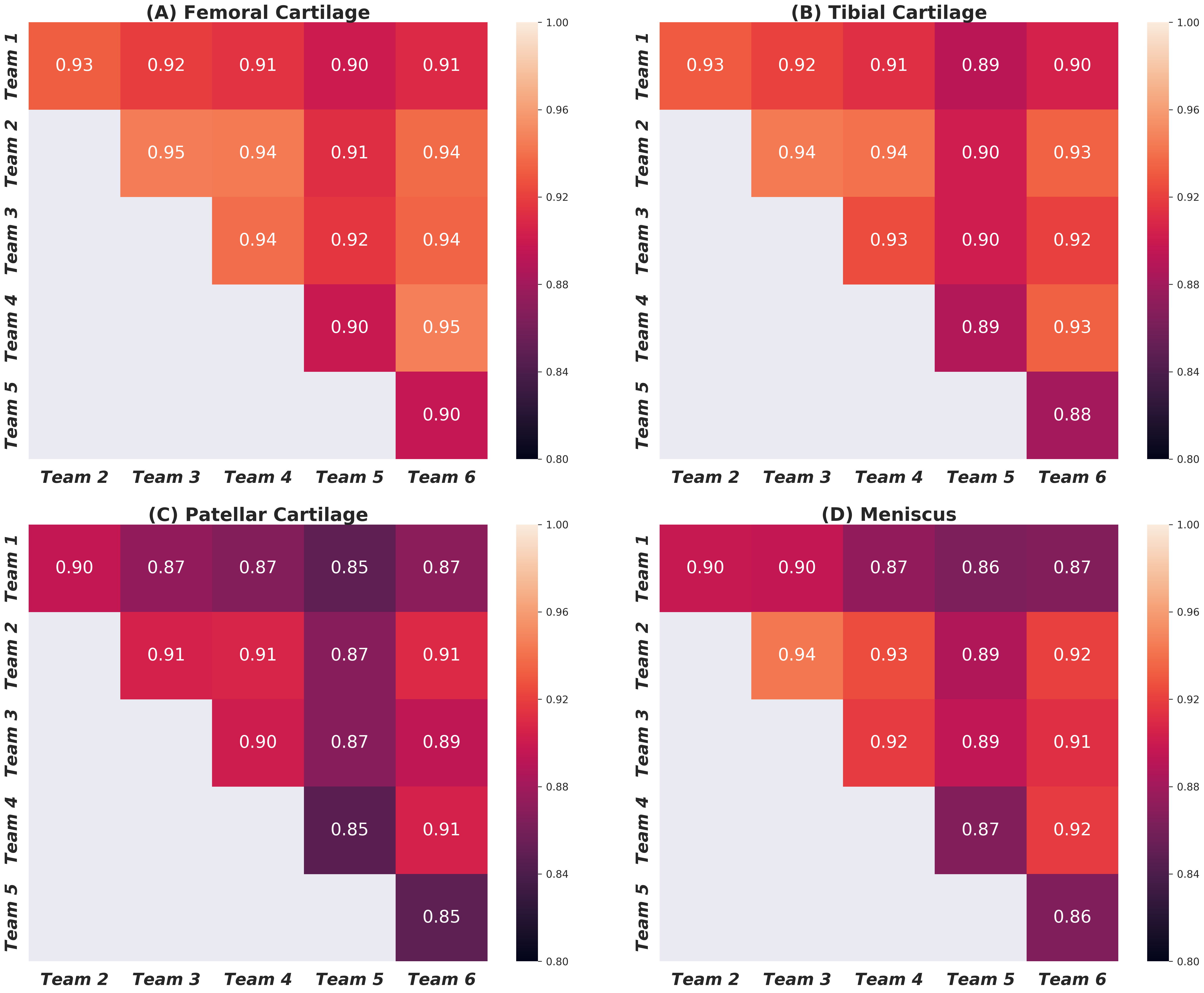}
  \caption{Dice correlations among segmentations from different networks for femoral cartilage (A), tibial cartilage (B), patellar cartilage (C), and meniscus (D). Strong correlation was observed for femoral cartilage, tibial cartilage, and menisci and moderately strong correlation was displayed for patellar cartilage.}
  \label{fig:dice}
\end{figure}

\subsection{Challenge Entries}
All networks achieved reasonably similar fidelity in segmenting all tissue structures as measured by standard segmentation metrics (Figure \ref{fig:perf-violin}A–D). Dice, VOE, root-mean-squared CV, and ASSD (mm) for all tissues ranged from 0.81–0.90, 0.17–0.31, 0.02–0.09, 0.18–0.40, respectively (Table \ref{tbl:metrics}). For femoral, tibial, and patellar cartilage, thickness errors ranged between 0.04mm–0.16mm. No significant differences were observed in Dice, CV, VOE, ASSD for femoral cartilage (p=1.0), tibial cartilage (p=1.0), patellar cartilage (p=1.0), and menisci (p=1.0) among the four top-performing networks ($T_2$, $T_3$, $T_4$, $T_6$). These four networks had significantly lower ASSD and higher Dice for femoral cartilage (p<0.05) in comparison to $T_1$ and $T_5$. Additionally, $T_2$, $T_3$, and $T_4$ had significantly higher Dice accuracy than $T_1$ and $T_5$ for tibial cartilage and meniscus (both p<0.05). No significant differences were observed in CV for femoral cartilage (p=0.14), tibial cartilage (p=0.17) and patellar cartilage (p=0.93). High variance in patellar cartilage segmentation was observed amongst all segmentation metrics, primarily due to one outlier subject (Table \ref{tbl:metrics}).

\begin{figure}
  \centering
  \includegraphics[width=8cm]{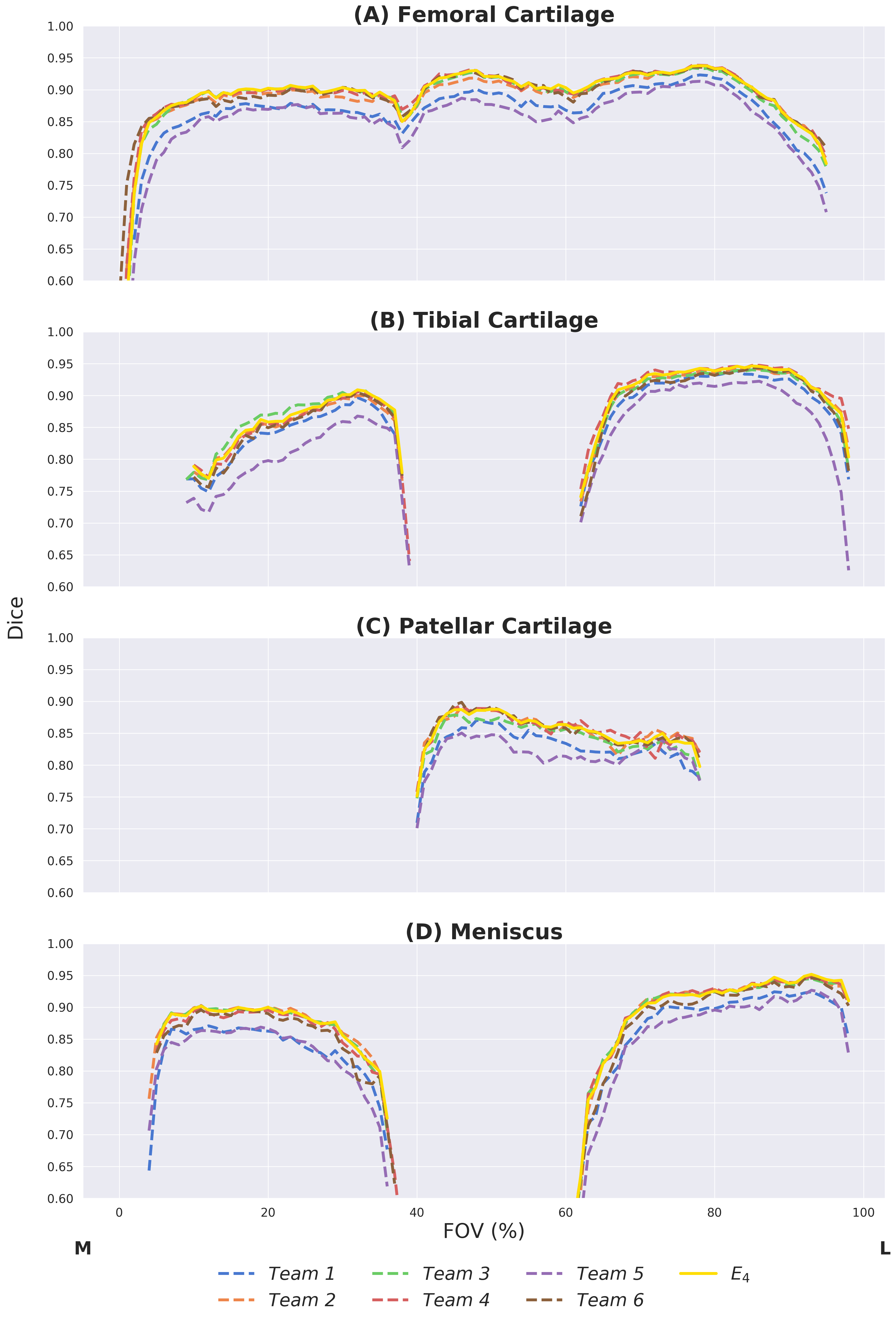}
  \caption{Depth-wise region of interest distribution (dROId). Segmentation accuracy using Dice as a function of slice location from the medial (M) to lateral (L) end. The field of view is normalized (0-100\%) based on the first and last slice with a ground truth segmentation in each scan. All networks have similar trends in performance across different regions of the knee. All networks share failure points at the intercondylar notch (\~40\% FOV) and have considerably lower performance in the medial condyle.}
  \label{fig:droid}
\end{figure}

Dice correlations between pairs of networks were $\rho_{Dice}$>0.90 for femoral cartilage, $\rho_{Dice}$>0.88 for tibial cartilage, $\rho_{Dice}$>0.86 for menisci, and $\rho_{Dice}$>0.85 for patellar cartilage (Figure \ref{fig:dice}). The top four networks demonstrated the strongest Dice correlations among femoral cartilage, tibial cartilage, and meniscus ($\rho_{Dice}$>0.94). Networks also displayed similar segmentation accuracy trends across DESS slices in the medial-lateral direction (Figure \ref{fig:droid}) and across KL-grades (Figure S1). All networks achieved higher Dice performance in the lateral condyle than in the medial condyle.

\begingroup
\begin{table}[]
 \caption{Average $\pm$ standard deviation segmentation performance for all submitted networks and majority-vote ensemble ($E_4$). Best performance among submitted networks for each metric per tissue is \textbf{bolded}. Results from majority-vote ensemble are \emph{italicized} if it achieves performance comparable or better than best performing submitted network. Coefficient of variation is calculated as root-mean-square value. VOE-- volumetric overlap error; RMS-CV--  root-mean-squared coefficient of variation; ASSD-- average symmetric surface distance.}
 \label{tbl:metrics}
 \centering
 \renewcommand{\arraystretch}{1}
 \setlength{\tabcolsep}{3pt}
 \begin{tabular}{ccccccccc}
  \toprule
  \multicolumn{2}{c}{} & \multicolumn{7}{c}{\thead{Networks}} \\
  \cmidrule(r){3-9}
  \thead{Tissue} & \thead{Metric} & \thead{\emph{Team 1}} & \thead{\emph{Team 2}} & \thead{\emph{Team 3}} & \thead{\emph{Team 4}} & \thead{\emph{Team 5}} & \thead{\emph{Team 6}} & \thead{$\mathbf{E_4}$} \\
  \midrule
  \multirow{5}{*}{\makecell{Femoral\\Cartilage}} & Dice & \dcell{0.88}{0.02} & \dcell{0.90}{0.02} & \dcell{0.90}{0.02} & \dcellbf{0.90}{0.02} & \dcell{0.87}{0.03} & \dcell{0.90}{0.02} & \dcellit{0.90}{0.02} \\
  & VOE & \dcell{0.22}{0.04} & \dcell{0.18}{0.03} & \dcell{0.18}{0.03} & \dcellbf{0.17}{0.03} & \dcell{0.23}{0.04} & \dcell{0.18}{0.03} & \dcellit{0.17}{0.03} \\
  & RMS-CV & \dcell{0.03}{0.02} & \dcell{0.03}{0.02} & \dcellbf{0.02}{0.01} & \dcell{0.02}{0.01} & \dcell{0.04}{0.03} & \dcell{0.02}{0.01} & \dcellit{0.02}{0.02} \\
  & ASSD (mm) & \dcell{0.23}{0.04} & \dcell{0.19}{0.03} & \dcell{0.18}{0.03} & \dcellbf{0.18}{0.03} & \dcell{0.24}{0.06} & \dcell{0.19}{0.05} & \dcellit{0.18}{0.03} \\
  & \makecell{Thickness\\Error(mm)} & \dcell{0.05}{0.05} & \dcell{0.05}{0.04} & \dcellbf{0.04}{0.03} & \dcell{0.06}{0.03} & \dcell{0.12}{0.07} & \dcell{0.11}{0.07} & \dcellit{0.04}{0.03} \\
  \midrule
  \multirow{5}{*}{\makecell{Tibial\\Cartilage}} & Dice & \dcell{0.87}{0.03} & \dcell{0.89}{0.03} & \dcellbf{0.89}{0.03} & \dcell{0.89}{0.04} & \dcell{0.85}{0.04} & \dcell{0.88}{0.03} & \dcellit{0.89}{0.03} \\
  & VOE & \dcell{0.23}{0.05} & \dcell{0.20}{0.05} & \dcellbf{0.20}{0.04} & \dcell{0.20}{0.06} & \dcell{0.26}{0.06} & \dcell{0.21}{0.05} & \dcellit{0.20}{0.05} \\
  & RMS-CV & \dcell{0.05}{0.03} & \dcell{0.05}{0.03} & \dcellbf{0.03}{0.02} & \dcell{0.06}{0.04} & \dcell{0.06}{0.04} & \dcell{0.05}{0.03} & \dcell{0.04}{0.03} \\
  & ASSD (mm) & \dcell{0.26}{0.14} & \dcell{0.25}{0.16} & \dcellbf{0.23}{0.14} & \dcell{0.25}{0.17} & \dcell{0.30}{0.18} & \dcell{0.29}{0.17} & \dcellit{0.23}{0.15} \\
  & \makecell{Thickness\\Error(mm)} & \dcell{0.06}{0.05} & \dcell{0.05}{0.03} & \dcell{0.04}{0.03} & \dcellbf{0.04}{0.03} & \dcell{0.14}{0.08} & \dcell{0.12}{0.07} & \dcell{0.05}{0.04} \\
  \midrule
  \multirow{5}{*}{\makecell{Patellar\\Cartilage}} & Dice & \dcell{0.83}{0.08} & \dcellbf{0.86}{0.07} & \dcell{0.85}{0.10} & \dcell{0.86}{0.07} & \dcell{0.81}{0.09} & \dcell{0.86}{0.07} & \dcellit{0.86}{0.08} \\
  & VOE & \dcell{0.29}{0.11} & \dcellbf{0.24}{0.09} & \dcell{0.25}{0.13} & \dcell{0.25}{0.10} & \dcell{0.31}{0.12} & \dcell{0.25}{0.10} & \dcellit{0.24}{0.11} \\
  & RMS-CV & \dcell{0.09}{0.07} & \dcellbf{0.06}{0.04} & \dcell{0.12}{0.10} & \dcell{0.08}{0.06} & \dcell{0.08}{0.06} & \dcell{0.07}{0.05} & \dcell{0.09}{0.08} \\
  & ASSD (mm) & \dcell{0.40}{0.46} & \dcell{0.23}{0.07} & \dcell{0.25}{0.14} & \dcellbf{0.23}{0.09} & \dcell{0.30}{0.12} & \dcell{0.34}{0.34} & \dcellit{0.23}{0.11} \\
  & \makecell{Thickness\\Error(mm)} & \dcell{0.10}{0.10} & \dcell{0.08}{0.06} & \dcellbf{0.06}{0.06} & \dcell{0.06}{0.05} & \dcell{0.16}{0.12} & \dcell{0.16}{0.11} & \dcellit{0.06}{0.04} \\ 
  \midrule
  \multirow{4}{*}{Meniscus} & Dice & \dcell{0.84}{0.03} & \dcellbf{0.88}{0.03} & \dcell{0.88}{0.03} & \dcell{0.87}{0.03} & \dcell{0.83}{0.03} & \dcell{0.87}{0.04} & \dcellit{0.88}{0.03}\\
  & VOE & \dcell{0.28}{0.04} & \dcellbf{0.22}{0.04} & \dcell{0.22}{0.04} & \dcell{0.22}{0.05} & \dcell{0.28}{0.05} & \dcell{0.23}{0.05} & \dcellit{0.22}{0.04} \\
  & RMS-CV & \dcell{0.05}{0.03} & \dcell{0.04}{0.02} & \dcellbf{0.03}{0.02} & \dcell{0.03}{0.02} & \dcell{0.06}{0.03} & \dcell{0.03}{0.02} & \dcell{0.05}{0.02} \\
  & ASSD (mm) & \dcell{0.40}{0.09} & \dcell{0.31}{0.08} & \dcellbf{0.30}{0.07} & \dcell{0.32}{0.10} & \dcell{0.38}{0.08} & \dcell{0.38}{0.19} & \dcellit{0.30}{0.08} \\
  \bottomrule
 \end{tabular}
\end{table}
\endgroup

\subsection{Cartilage Thickness}
\begin{figure}
  \centering
  \includegraphics[width=12cm]{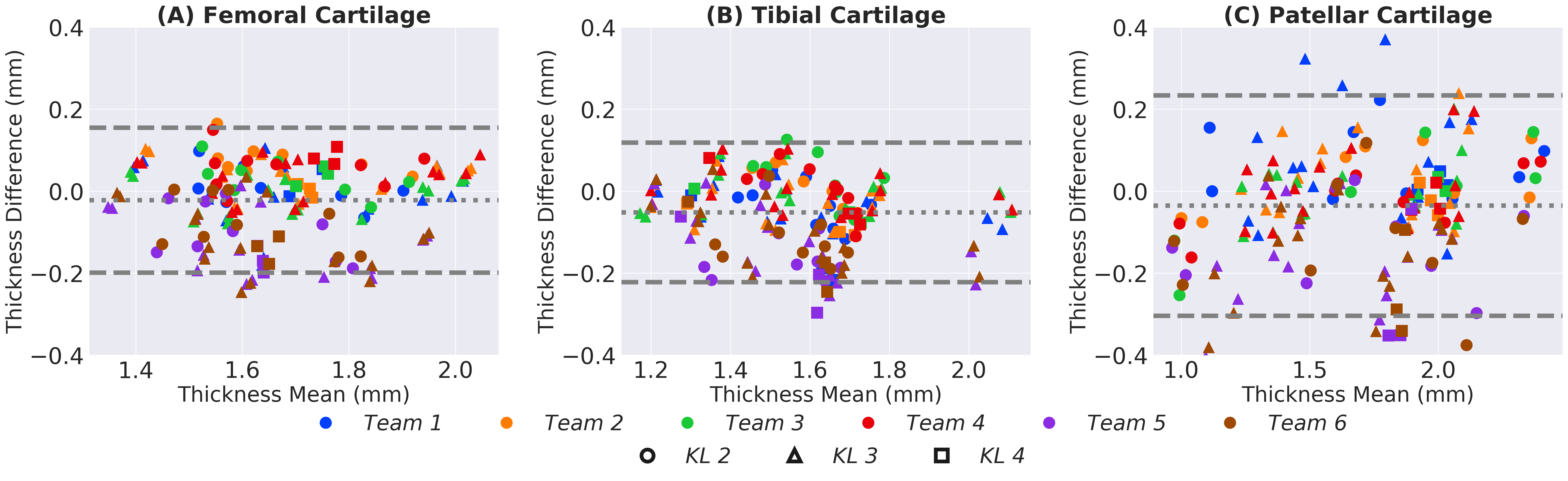}
  \caption{Thickness (per scan, Kellgren-Lawrence: 1-4) compared between network segmentations and ground truth segmentations for femoral cartilage (A), tibial cartilage (B), and patellar cartilage (C). Negligible bias (gray dotted line) was observed for all three tissues among all networks. 2*SEM confidence interval (between gray dashed lines) was broad, indicating high variance in thickness difference. Thickness difference was computed as the difference between thickness calculated on network (predicted) segmentation and thickness calculated from ground truth segmentation.}
  \label{fig:thickness}
\end{figure}

\begin{figure}
  \centering
  \includegraphics[width=12cm]{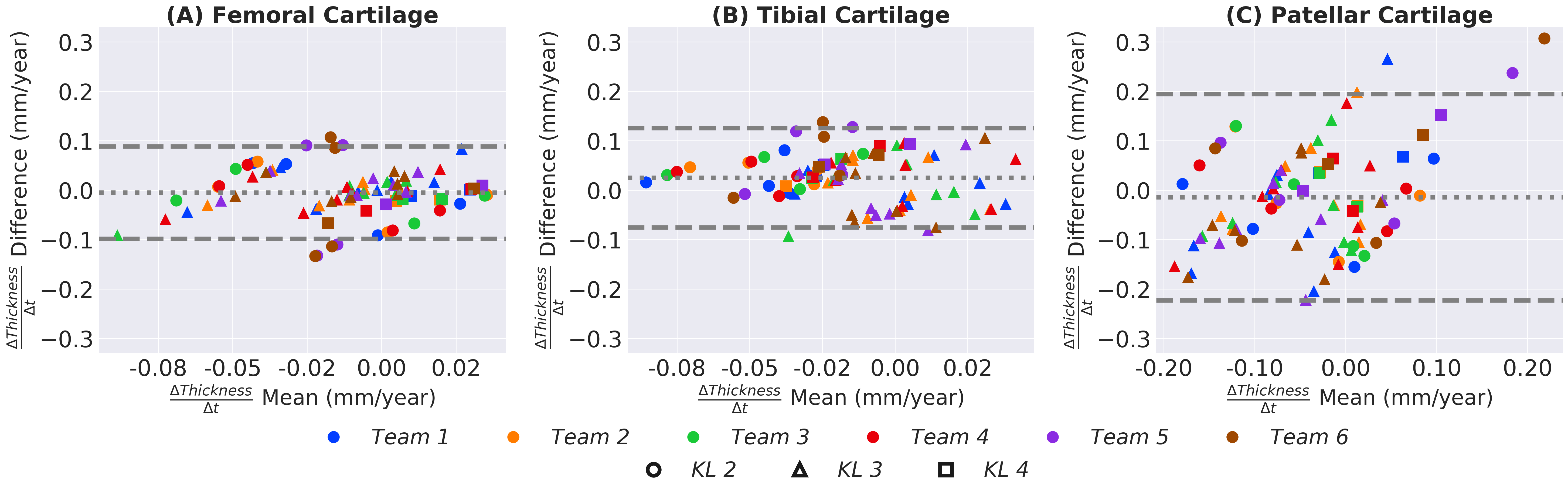}
  \caption{Longitudinal thickness change (per subject, Kellgren-Lawrence at timepoint 1: 2-4) measured by network segmentations and ground truth segmentations for femoral cartilage (A), tibial cartilage (B), and patellar cartilage (C). Positive difference values (y-axis) indicate overestimation of longitudinal thickness change.  Negligible bias (gray dotted line) was observed for all tissues. 2*SEM confidence interval (between gray dashed lines) is relatively small for femoral cartilage and tibial cartilage, indicating better longitudinal estimates. Networks appear to randomly underestimate and overestimate thickness changes.}
  \label{fig:thickness-longitudinal}
\end{figure}

Thickness errors from segmentations from $T_5$ and $T_6$ were significantly worse than those achieved by other networks for all three cartilage surfaces (Figure \ref{fig:perf-violin}E) (p<0.05) but were negligible among the other four networks (p=0.99). Among these four networks, median percent error in thickness estimates was ~5\% for femoral and tibial cartilage. Median percent error in patellar cartilage thickness estimates was larger and more variable (Figure S2). There was minimal systematic underestimation of cartilage thickness (femoral cartilage: -0.02mm, tibial cartilage: -0.05mm, patellar cartilage: -0.03mm) (Figure \ref{fig:thickness}). Bland-Altman limits of agreement 95\% confidence intervals were 0.35mm for femoral cartilage, 0.34mm for tibial cartilage, and 0.54mm for patellar cartilage. Additionally, there was minimal bias in estimated longitudinal thickness changes between segmentations across all networks and ground-truth masks (femoral cartilage: 0.00mm, tibial cartilage: 0.03mm, patellar cartilage: -0.01mm) (Figure \ref{fig:thickness-longitudinal}). Bland-Altman limits of agreement 95\% confidence intervals for longitudinal thickness changes were 0.19mm, 0.20mm, and 0.42mm for femoral cartilage, tibial cartilage, and patellar cartilage, respectively.

\begin{figure}
  \centering
  \includegraphics[width=12cm]{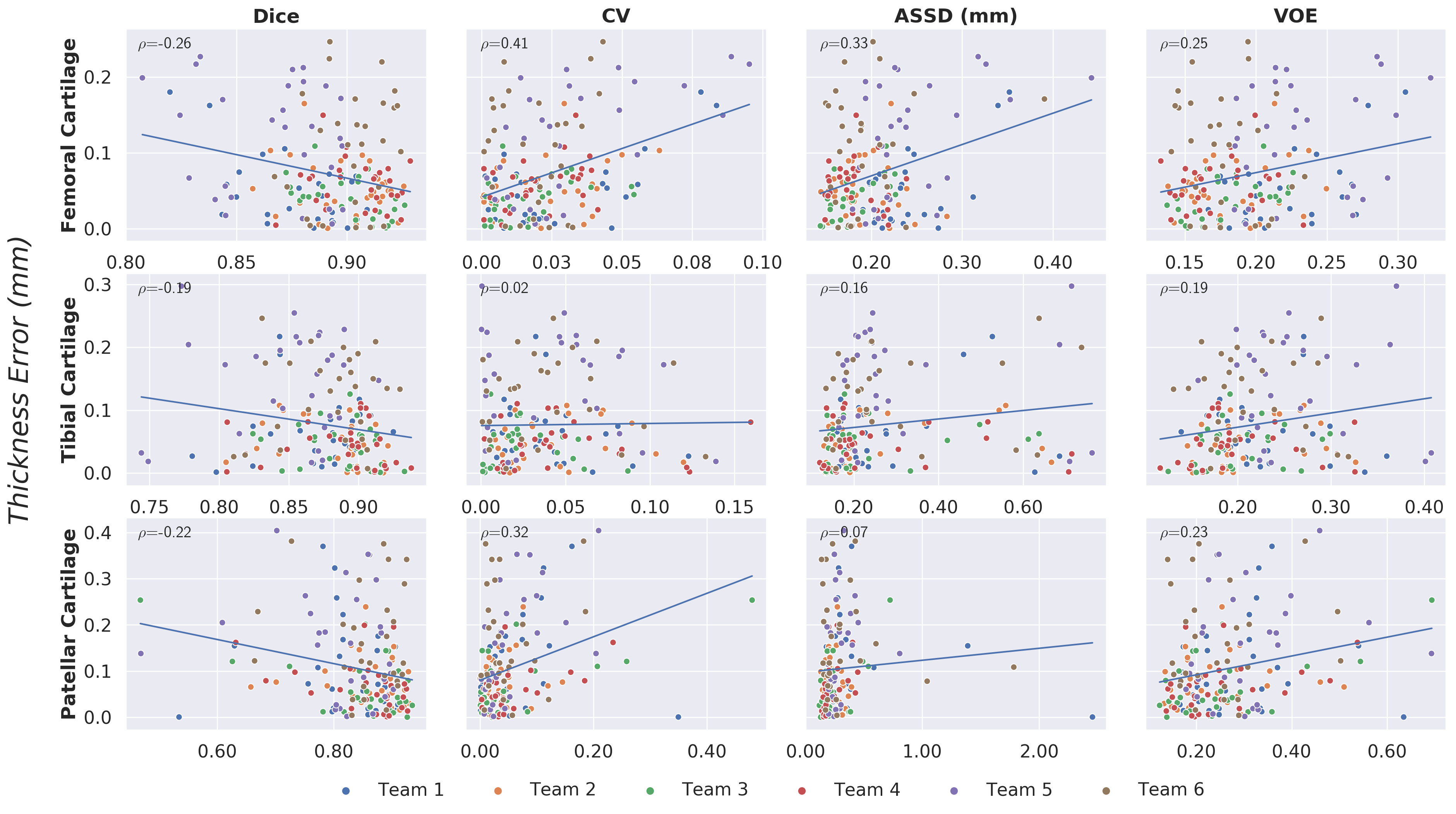}
  \caption{Correlation between pixel-wise segmentation metrics and cartilage thickness error as measured by the Pearson’s correlation coefficient ($\rho$). Minimal correlation was observed for all tissues across networks, all of which achieved high segmentation performance. This may suggest that given high performance among models as measured by pixel-wise segmentation metrics, there is a negligible difference in diagnostic metrics.}
  \label{fig:metrics-correlation}
\end{figure}

There was low correlation between pixel-wise segmentation accuracy metrics and cartilage thickness ranged from very-weak to moderate (highest Pearson-R=0.41). Highest correlations between all pixel-wise metrics and cartilage thickness were observed with femoral cartilage thickness (Pearson-R>0.25), while very-weak correlation among these metrics was observed with tibial cartilage (Pearson-R<0.2) (Figure \ref{fig:metrics-correlation}). CV had the highest correlation with femoral and patellar cartilage thickness (Pearson-R=0.41, 0.32).

\subsection{Ensemble Comparison}
The majority-vote ensemble ($E_4$) achieved similar performance to the submitted networks for both pixel-wise and thickness metrics (Table \ref{tbl:metrics}). No significant performance difference was observed between the ensemble and the networks with the best performance for quantitative segmentation metrics ($T_2$, $T_3$, $T_4$, $T_6$) as well as networks with the best performance for cartilage thickness ($T_1$, $T_2$, $T_3$, $T_4$) (p=1.0). The ensemble also displayed similar segmentation accuracy trends across DESS slices in the medial-lateral direction as the individual networks (Figure \ref{fig:droid}). Both optimal upper-bound ensemble networks ($E_+^\ast$ and $E_-^\ast$) performed significantly better than the majority-vote ensemble in pixel-wise metrics (p<0.05) but not thickness error (p=1.0). No significant performance difference was observed between the two upper-bound ensemble networks for all tissues (p=0.7) (Table \ref{tbl:opt-ensemble-metrics}). Dice correlations between segmentations from these two ensembles were 0.96, 0.96, 0.95, and 0.95 for femoral cartilage, tibial cartilage, patellar cartilage, and meniscus, respectively (Table \ref{tbl:opt-ensemble-metrics}).

\section{Discussion}
In this study, we organized a knee MRI segmentation challenge consisting of a common MRI sequence that can be deployed in prospective studies, which contains information sensitive to soft-tissue degeneration in OA. We organized the OAI segmentation data and standardized dataset splits with balanced demographics in an easy-to-use format for future research studies. We developed a framework to compare and evaluate the performance of challenge submission entries for segmenting articular cartilage and meniscus on a standardized dataset split. We showed that CNNs achieved similar performance independent of network architecture and training design for segmenting all tissues. In addition, we evaluated differences in network performance on both segmentation and osteoarthritis progression metrics. Although a high segmentation accuracy was achieved by all models and ensembles, only a weak correlation between segmentation accuracy metrics and cartilage thickness was observed. Moreover, we explored how a network ensemble approach can be a viable technique for combining outputs from multiple high-performance networks and how simulated combinations of network outputs can be used to quantify performance bounds and error profiles for ensembles.

\subsection{Designing Networks}
Despite the vast variety of network approaches, most methods achieved similar segmentation and thickness accuracy across all tissues. While some networks had significantly lower performance compared to submissions for other teams, all networks shared a high Dice correlation, suggesting strong concordance in volumetric similarity of the segmentations. In addition, near-identical slice-wise Dice accuracies and failure regions indicated that all networks systematically performed worse in the intercondylar notch and the medial compartment, which is more commonly affected in subjects with OA \cite{dieppe1998osteoarthritis}. The similarity in performance and limitations may suggest that independent networks, regardless of their design and training framework, learn to represent and segment the knee in similar ways. It may also indicate that state-of-the-art networks may primarily be data-limited and may reap minimal benefit from architecture and/or training protocol optimizations.

Due to a similarity in learned image representations, the $E_4$ voting ensemble performed similarly to the individual networks. When errors among models are minimally correlated, majority voting ensembles can improve performance over best-performing individual networks. The minimal performance gain from $E_4$ may indicate the high correlation among errors from individual networks, which was also observed in the high dice correlations among network segmentations. This may also suggest that individual, high-accuracy knee segmentation models can achieve similar performance to their ensemble counterparts. In the case of lower segmentation and thickness accuracy models, the voting ensemble improved performance across all metrics, even though independently different sets of networks perform poorly across different metric types.

Moreover, empirically gauging performance of different models in the wild is difficult, as it requires acquiring ground truth labels for comparison. In a prospective deployment of segmentation models without the availability of ground-truth labels, ensembles can provide implicit regularization by limiting the impact of any arbitrary poorly performing network on the output. Furthermore, while voting ensembles may be an effectively method for regularizing network outputs in the wild, training ensembles to learn relative spatial weightings among models may be a more exhaustive method for improving overall performance. In such cases, high-accuracy models with low concordance in errors could provide highly complementary information that may be useful in training ensembles.

The ensemble upper-bound computation may be helpful in empirically quantifying the extent to which ensembles can leverage variations in segmentation. While trivial segmentations (all 0s or all 1s) would saturate the upper-bound, none of the six models presented trivial solutions. The performance between the optimized upper-bound ensembles $E_+^\ast$ and $E_-^\ast$ was highly concordant in both pixel-wise and thickness metrics. $E_+^\ast$ and $E_-^\ast$ isolated errors in the segmentation to false negatives and false positives, respectively. This may indicate that the incidence rate of false negatives and false positives is well balanced and that either error has an equal contribution to the overall error. This may suggest a reduced need to artificially weight networks to handle class imbalance in knee segmentation tasks, such as that through the Tversky loss function.

\begingroup
\begin{table}[]
 \caption{Average $\pm$ standard deviation segmentation performance for and dice correlation ($\rho_{Dice}$) between optimal true positive ($E_+^\ast$) and optimal true negative ($E_-^\ast$) ensembles. There was no significant difference observed between the two networks across all metrics and high dice correlations for all tissues. Value for coefficient of variation (CV) is calculated as root-mean-square (RMS), not average.}
 \label{tbl:opt-ensemble-metrics}
 \centering
 \renewcommand{\arraystretch}{1}
 \setlength{\tabcolsep}{8pt}
 \begin{tabular}{cccc}
  \toprule
  \multicolumn{2}{c}{} & \multicolumn{2}{c}{\thead{Networks}} \\
  \cmidrule(r){3-4}
  \thead{Tissue} & \thead{Metric} & \thead{$\mathbf{E_+^\ast}$} & \thead{$\mathbf{E_-^\ast}$} \\
  \midrule
  \multirow{6}{*}{\makecell{Femoral\\Cartilage}} & Dice & \dcell{0.98}{0.01} & \dcell{0.98}{0.01} \\
  & VOE & \dcell{0.04}{0.01} & \dcell{0.04}{0.01}  \\
  & RMS-CV & \dcell{0.02}{0.01} & \dcell{0.02}{0.01}  \\
  & ASSD (mm) & \dcell{0.03}{0.01} & \dcell{0.04}{0.01} \\
  & Thickness Error(mm) & \dcell{0.04}{0.01} & \dcell{0.04}{0.02}  \\
  & $\rho_{Dice}$ & \multicolumn{2}{c}{\dcell{0.96}{0.01}} \\
  \midrule
  \multirow{6}{*}{\makecell{Tibial\\Cartilage}} & Dice & \dcell{0.98}{0.01} & \dcell{0.98}{0.01} \\
  & VOE & \dcell{0.04}{0.02} & \dcell{0.04}{0.03}  \\
  & RMS-CV & \dcell{0.03}{0.01} & \dcell{0.03}{0.01}  \\
  & ASSD (mm) & \dcell{0.03}{0.03} & \dcell{0.07}{0.09}  \\
  & Thickness Error (mm) & \dcell{0.03}{0.02} & \dcell{0.03}{0.02}  \\
  & $\rho_{Dice}$ & \multicolumn{2}{c}{\dcell{0.96}{0.01}} \\
  \midrule
  \multirow{6}{*}{\makecell{Patellar\\Cartilage}} & Dice & \dcell{0.97}{0.03} & \dcell{0.98}{0.01} \\
  & VOE & \dcell{0.06}{0.05} & \dcell{0.04}{0.03}  \\
  & RMS-CV & \dcell{0.04}{0.03} & \dcell{0.02}{0.01}  \\
  & ASSD (mm) & \dcell{0.04}{0.02} & \dcell{0.04}{0.03}  \\
  & Thickness Error (mm) & \dcell{0.05}{0.03} & \dcell{0.04}{0.03}  \\
  & $\rho_{Dice}$ & \multicolumn{2}{c}{\dcell{0.95}{0.02}} \\
  \midrule
  \multirow{5}{*}{Meniscus} & Dice & \dcell{0.97}{0.01} & \dcell{0.98}{0.01} \\
  & VOE & \dcell{0.06}{0.02} & \dcell{0.03}{0.02}  \\
  & RMS-CV & \dcell{0.03}{0.01} & \dcell{0.02}{0.01}  \\
  & ASSD (mm) & \dcell{0.06}{0.03} & \dcell{0.04}{0.03}  \\
  & $\rho_{Dice}$ & \multicolumn{2}{c}{\dcell{0.95}{0.01}} \\
  \bottomrule
 \end{tabular}
\end{table}
\endgroup

\subsection{Segmentation Metrics and Thickness}
Networks performed similarly among segmentation metrics, including thickness measures. Median cartilage thickness errors were <0.2mm, roughly half the resolution of the DESS voxels among networks with high volumetric (VOE, ASSD) and overlap (Dice) performances. The magnitude of thickness errors was also slightly below observed 1-year changes in cartilage thickness (~0.2-0.3 mm) \cite{maschek2014rates, cotofana2014loss, wirth2011comparison}. While thickness errors are both sub-scan resolution and within observed change, additional work towards reducing the high variability of these estimates may help make these networks more viable for clinical use.

Compared to cross-sectional thickness estimates, longitudinal thickness estimates had considerably lower variability, which may indicate that all networks systematically overestimate or underestimate cartilage thickness per subject. In the case of longitudinal estimates, this per-subject systematic bias may be helpful as it reduces the error when evaluating longitudinal cartilage thickness changes. The consistency in estimation over time may suggest that these networks capture anatomical features that change minimally-over time. The relative time-invariance of these features may be useful in fine-tuning subject-specific networks on individual subject scans to perform more robustly on future scans of that subject.

The larger variance in thickness estimates compared to pixel-wise metrics was further indicated by the stark difference in performance amongst the two types of metrics. Individual networks that performed best among pixel-wise metrics ($T_2$, $T_3$, $T_4$, $T_6$) did not necessarily perform the best in estimating thickness and vice versa. Even between optimal upper-bound ensembles ($E_+^\ast$ and $E_-^\ast$)  and the majority vote ensemble ($E_+^\ast$), where there was a significant different in most segmentation metrics, there was no difference in the error of thickness estimates between the upper-bound and majority-vote ensembles.

Overall, the correlations between standard segmentation metrics and cartilage thickness were weak. These factors may suggest that using traditional evaluation metrics on high-performing models may not be predictive of differences in thickness accuracy outcomes among high-performance models. Small improvements in segmentation metrics among these models may not correspond to increased thickness accuracy. It may also indicate that information learned from pixel-level segmentation accuracy and tissue-level thickness accuracy metrics are somewhat complementary among these models. A loss function designed to optimize for a combination of segmentation and thickness accuracy may regularize model performance among both segmentation and clinical endpoints.

\subsection{Code and Weights Availability}
Training, validation, and testing partitions are available in the downloadable csv document (supplement). Code and learned model weights for $T_3$\footnote{\url{http://github.com/perslev/MultiPlanarUNet}}, $T_5$\footnote{\url{http://github.com/ali-mor/IWOAI_challenge}}, and $T_6$\footnote{\url{http://github.com/ad12/DOSMA}} are also available.

\subsection{Limitations}
Despite the large set of networks, there were certain limitations in the study. First, all compared methods leveraged CNNs with minimal post-processing. Additional non-deep-learning approaches and practical considerations for post-processing, such as conditional random smoothing fields, can be explored to further refine CNN outputs. Additionally, while the majority-vote ensemble simulated practical methods for combining outputs from different models with limited access to models, ensemble learning from model logits may improve accuracy by learning relative weighting among different models. However, this would require access to model outputs from training data in addition to the testing data, which is challenging in the context of multi-institutional studies and will be the focus of future studies.

\section{Conclusion}
In this study, we standardized a dataset partition using an MRI sequence that can be prospectively deployed to train and evaluate knee segmentation algorithms. We established a generalized framework for interpreting the clinical utility of such segmentations beyond using longstanding segmentation metrics. Using deep-learning-based segmentation algorithms from multiple institutions, we showed that networks with varying training paradigms achieved similar performance. Through these multiple networks, we demonstrated the efficacy of using majority-vote ensembles in cases of limited access to training resources or explicit network parameters. Moreover, among individual models achieving high segmentation performance, segmentation accuracy metrics were weakly correlated with cartilage thickness endpoints.

\subsubsection*{Grant Support}
Contract grant sponsor: National Institutes of Health (NIH); contract grant numbers NIH R01 AR063643, R01 EB002524, K24 AR062068, and P41 EB015891, R00AR070902, R61AR073552, R01 AR074453. Contract grant sponsor: National Science Foundation (DGE 1656518). Contract grant sponsor: GE Healthcare and Philips (research support). Contract grant sponsor: Stanford University Department of Radiology Precision Health and Integrated Diagnostics Seed Grant. Image data was acquired from the Osteoarthritis Initiative (OAI). The OAI is a public-private partnership comprised of five contracts (N01-AR-2-2258; N01-AR-2-2259; N01-AR-2-2260; N01-AR-2-2261; N01-AR-2-2262) funded by the National Institutes of Health, a branch of the Department of Health and Human Services, and conducted by the OAI Study Investigators. Private funding partners include Merck Research Laboratories; Novartis Pharmaceuticals Corporation, GlaxoSmithKline; and Pfizer, Inc. Private sector funding for the OAI is managed by the Foundation for the National Institutes of Health. This manuscript was prepared using an OAI public use data set and does not necessarily reflect the opinions or views of the OAI investigators, the NIH, or the private funding partners.

\subsubsection*{Disclosures}
G.G. and B.H. have received research support from GE Healthcare and Philips. V.P. and S.M. have received research support from GE Healthcare. B.H. is a shareholder of LVIS Corporation. A.C. has provided consulting services to SkopeMR, Inc., Subtle Medical, Chondrometrics GmbH, Image Analysis Group, Culvert Engineering, and Edge Analysis and is a shareholder of Subtle Medical, LVIS Corporation, and Brain Key. E.D is a shareholder of Biomediq and Cerebriu. A.P is a shareholder of Cerebriu. None of the mentioned organizations were involved in the design, execution, data analysis, or the reporting of this study.

\small
\bibliography{ms}

\end{document}

\maketitle

\section*{Ensemble Upper Bounds}
To empirically evaluate the upper bound of segmentation sensitivity and specificity that can be achieved by a combination of models, we compute two upper-bound performance ensembles, optimized for either true positives or true negatives, respectively. The performance ensemble optimized for true positives ($E_+^\ast$) generates masks for tissues by consolidating only true positives from masks generated by each of the subsidiary models. Similarly, the performance ensemble optimized for true negatives ($E_-^\ast$) generates masks by consolidating only true negatives from masks of subsidiary models. As both ensembles only consolidate accurate labels among all models, any ensemble of the ordinal individual model outputs cannot exceed the performance bounds established by these combinations. Combinations were generated from binary segmentation masks produced by models $T_1$-$–T_6$.

Because $E_+^\ast$ only consolidates true positive pixels, it lacks false positives, rendering all errors from false negative labels. Conversely, $E_-^\ast$ lacks false negatives, making all errors a consequence of false positive labels. The relative performance of the two networks suggests the extent to which all networks are insensitive to areas where the tissue was annotated and overly sensitive to areas where the tissue was not annotated.

\section*{Figures}
\begin{figure}[H]
\renewcommand\thefigure{S1} 
\centering
\includegraphics[width=14cm]{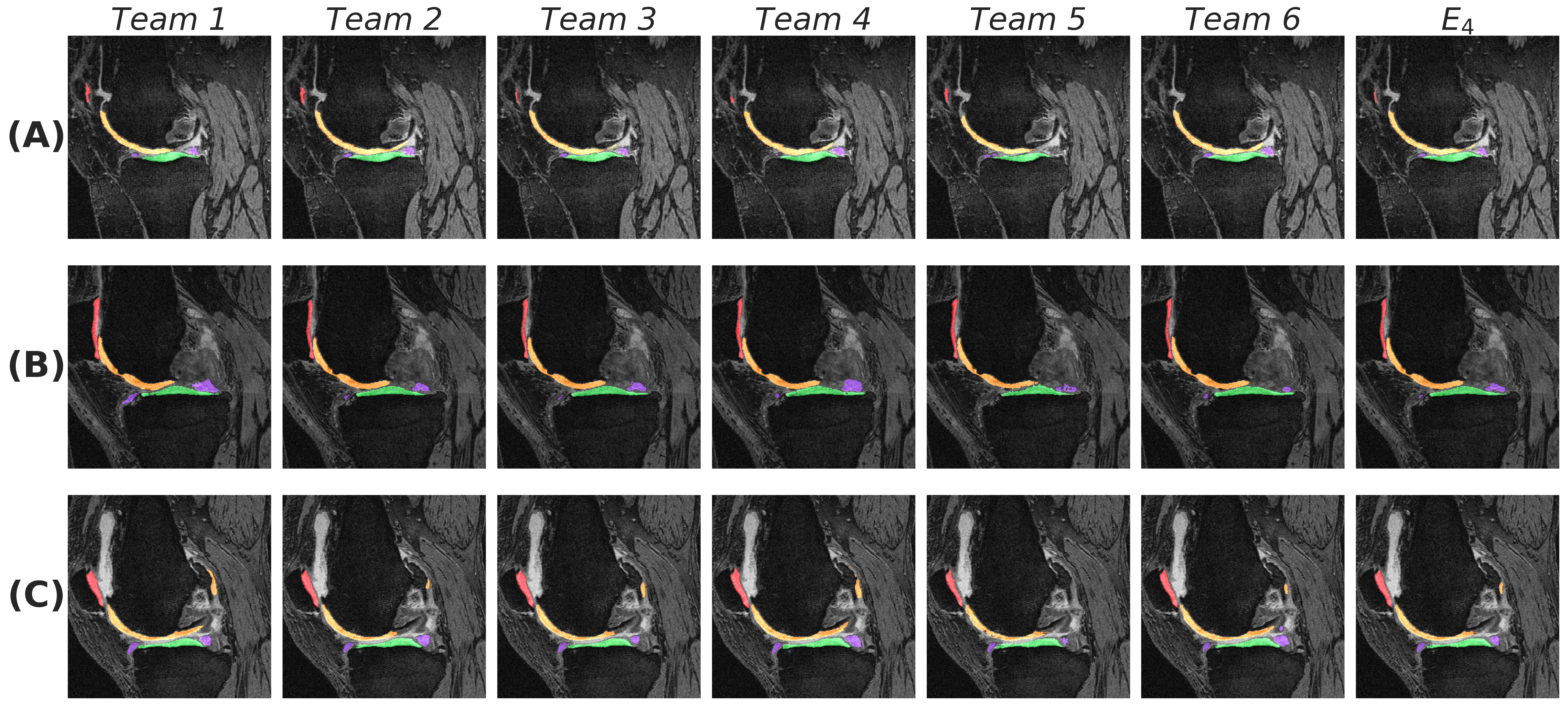}
\caption{Sample segmentations (1.25$\times$ center zoom) of the lateral condyle in subjects of KL grades 2-4 (A-C, respectively). Tissue (color): femoral cartilage (orange), tibial cartilage (green), patellar cartilage (red), meniscus (purple). Segmentation differences appeared negligible among all networks, including the majority-vote ensemble ($E_4$).}
\label{supp-fig:sample-segmentations}
\end{figure}

\begin{figure}[H]
\renewcommand\thefigure{S2} 
\centering
\includegraphics[width=14cm]{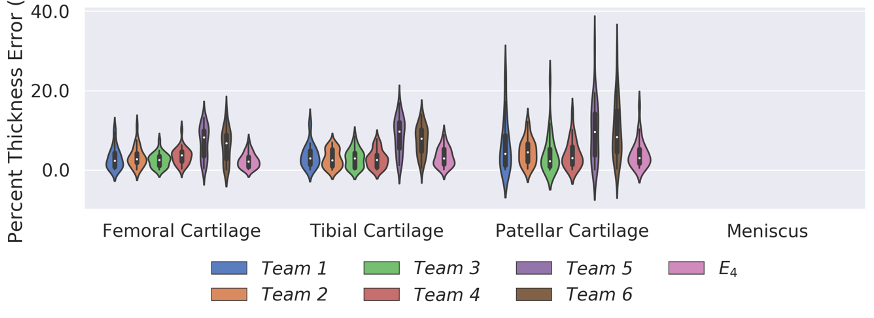}
\caption{Percent thickness error of networks submitted to segmentation challenge and majority-vote ensemble ($E_4$) for cartilage structures. Networks performing well on thickness metrics (Team 1, Team 2, Team 3, Team 4) achieved a median percent thickness error of ~5\% for femoral and tibial cartilage. However, performance on patellar cartilage was considerably worse and had more variable performance (longer tails).}
\label{supp-fig:percent_thickness_error}
\end{figure}